\documentstyle[psfig]{mn}
\def\H0{{\it H}$_0$}
\def\Ms{{\it M}$_\odot$}

\def\q0{{\it q}$_0$}
\def\kmps{km~s$^{-1}$}

\def\ergps{erg~s$^{-1}$}

\def\Ms{{\it M}$_\odot$}

\def\nH{$N_{\rm H}$\thinspace} 

\def\psqcm{cm$^{-2}$}
\def\ergpspsqcm{erg~cm$^{-2}$~s$^{-1}$}

\def\cps{ct\thinspace s$^{-1}$}

\def\pcubcm{cm$^{-3}$}
\def\ergcmps{erg\thinspace cm\thinspace s$^{-1}$}
\def\nW{$N_{\rm W}$}

\title[X-ray emission from NGC4395] 
{X-ray absorption and rapid variability
of the dwarf Seyfert nucleus of NGC4395} 
\author[K. Iwasawa et al] 
{\parbox[]{6.5in} {K.~Iwasawa$^1$, A.C.~Fabian$^1$, O.~Almaini$^2$, P.~Lira$^{2,3}$, A.~Lawrence$^2$, K.~Hayashida$^4$ and H.~Inoue$^5$}\\
\\
$^1$Institute of Astronomy, Madingley Road, Cambridge CB3 0HA\\ 
$^2$Institute for Astronomy, University of Edinburgh, Royal Observatory, Blackford Hill, Edinburgh EH9 3HJ\\
$^3$Department of Physcs and Astronomy, University of Leicester, University Road, Leicester, LE1 7RH\\
$^4$Department of Earth and Space Science, Osaka University, Machikaneyama, Toyonaka, Osaka 560-0043, Japan\\
$^5$Institute of Space and Astronautical Science, Yoshinodai, Sagamihara, Kanagawa 229-8510, Japan\\
}
\date{}

\begin{document}

\maketitle

\begin{abstract}
We report the detection of an absorbed central X-ray source and its strong,
rapid, variability in NGC4395, the least luminous Seyfert nucleus
known.  The X-ray source exhibits a number of flares with factors of
3--4 flux changes during a half day ASCA observation.  The shortest
doubling time observed is about 100 s.  Such X-ray variability is in
constrast to the behaviour of other low luminosity active galaxies and 
resembles that of higher luminosity Seyfert 1 galaxies. It provides further
support for an accreting black hole model rather than an extreme
stellar process in accounting for the nuclear activity of NGC4395.
The ASCA spectrum shows a power-law continuum of photon-index $\Gamma
= 1.7\pm 0.3$ with a Fe K line marginally detected at $\sim 6.4$ keV.
The soft X-ray emission below 3 keV is strongly attenuated by absorption.  
The energy spectrum in this absorption band shows a dramatic change
in response to the variation in continuum luminosity.
A variable warm absorber appears to
be the most likely explanation to account for the spectral change.
The absorption-corrected 2--10 keV luminosity is
$4\times 10^{39}$\ergps\ for a source distance of 2.6 Mpc, 
and at 1 keV is one order of magnitude above previous ROSAT
estimates, which affects the appearance of the wide-band
spectral energy distribution and photoionization calculations.  The
rapid X-ray variation is consistent with a black hole of a few times $10^4$
\Ms, as suggested by the optical results and the small bulge of this dwarf 
galaxy.
Such a light black hole is favoured also in order for 
the Eddington ratio ($L_{\rm Bol}/L_{\rm Edd}$) to be above the range of 
ADAFs, which would clearly fail to explain the observed X-ray variability.
The nuclear source of NGC4395 is therefore consistent with
a scaled-down version of 
higher luminosity Seyfert nuclei, with an intermediate mass 
($10^4$--$10^5$\Ms) black hole,
unlike the nearby low luminosity active galaxies 
in which underfed massive black holes are suspected to reside.
\end{abstract}


\section{introduction}

Recent kinematical studies of nearby galaxies have shown
that massive dark objects (MDOs) appear to be ubiquitous 
at the centre of galaxies (e.g, Kormendy \& Richstone 1995),
and that most luminous galaxies seem to have MDOs, most likely black holes,
with mass ranging $10^6$--$10^9$\Ms\ (e.g., Richstone et al 1998).
Many of these objects are, however, significantly underluminous relative to
the estimated gas supply (e.g, Fabian \& Rees 1995).
To account for this, the Advection-Dominated Accretion Flow (ADAF) 
solutions relevant for low accretion rates have been proposed.

Recent extensive optical spectroscopic surveys (e.g., Ho, Filippenko \&
Sargent 1997a) have revealed 
that a fair fraction of nearby galaxies exhibit some level of activity
in their nuclei. 
Their optical spectra typically show LINER (Heckman 1980) properties
and broad H$\alpha $ emission is seen in some of them 
(Ho, Filippenko \& Sargent 1997b).
Whether these low activity objects are scaled-down versions of 
more luminous Seyfert 1 and QSO nuclei or an alternative mechanism
is responsible is an important issue.

Such low luminosity active galactic nuclei (dwarf AGN) 
tend to show weak X-ray variability
compared with the higher luminosity
Seyfert 1 galaxies investigated by Nandra et al (1997).
Ptak et al (1998) interpreted this as evidence for ADAFs operating 
at low accretion rate in dwarf AGN.

NGC4395 hosts one of the dwarf Seyfert nuclei in the Ho et al
(1997a,b) sample, and the least luminous AGN known.  This dwarf galaxy
is a late-type spiral of low surface brightness with no significant
bulge. A study of stellar kinematics indicates a shallow gravitational
potential of the small bulge ($<8\times 10^4$\Ms, Filippenko \& Ho
2000) and hence the central black hole (e.g., Magorrian et al
1998). Similarly small black hole masses ($\sim 10^5$\Ms) have been
estimated from optical investigations of the active nucleus (Lira et
al 1999; Kraemer et al 1999).

A point-like optical nucleus located in the centre of the 
galaxy shows emission-line properties more resembling a Seyfert 1
nucleus than a LINER (Filippenko \& Sargent 1989; Filippenko, Ho \&
Sargent 1993).  Ho et al (1997a) classified NGC4395 as a Seyfert 1.8
on account of the presence of broad permitted line emission
(FWZI(H$\alpha)\sim 5000$\kmps) and high excitation condition.  A
number of coronal lines like [FeVII]$\lambda 6087$, [FeX]$\lambda
6374$ (e.g., Ho et al 1997b; Kraemer et al 1999) are detected and the
broad Balmer emission was found to be variable (Lira et al 1999).  The
contribution of stellar light to the nuclear spectrum appears to be
minimal as no significant absorption lines are seen in the HST UV
spectrum (Filippenko, Ho \& Sargent 1993), although weak CaIIK
absorption was found by Lira et al (1999) who estimate the stellar
light contribution to be about 10 per cent in the blue band.  The
apparent deficit of ionizing photons relative to the observed H$\beta$
luminosity (e.g., Moran et al 1999), similar to some Seyfert 2 nuclei,
indicates that the UV continuum source is attenuated by some
obscuration in the line of sight, while the narrow-line region (NLR)
seems to be little obscured apart from Galactic extinction
($E(B-V)=0.017$, Kraemer et al 1999).  Electron scattering is
suggested as an origin of the optical continuum polarisation (6.7 per
cent) reported by Barth, Filippenko \& Moran (1999), but the result is
also consistent with transmission through aligned dust.

NGC4395 has been observed in X-rays with the ROSAT PSPC and HRI. 
Lira et al (1999) and Moran et al (1999) independently analyzed 
the data and found the nuclear X-ray source to vary by a factor of 
$\sim 2$ in two weeks. The soft X-ray luminosity is estimated to be
$10^{38}$\ergps, which led them to interpret the nuclear source
as X-ray quiet compared with its wide band spectral energy distribution.
We observed NGC4395 in the higher energy X-ray band with ASCA and find that
the soft X-ray emission observed with ROSAT is faint due to absorption
and the primary X-ray source has a luminosity one order of magnitude
above the ROSAT estimate, when corrected for the absorption.
We also find the X-ray source to be extremely variable unlike the 
other dwarf AGN studied by Ptak et al (1998).
The properties of the absorber and the central source, assuming
an intermediate mass black hole, are discussed on the basis of
the X-ray results.

\section{Observations and data reduction}

\begin{table}
\begin{center}
\caption{The mean count rates of NGC4395 observed with the ASCA four 
detectors. Values for large ($\phi_{\rm Large}$)
and small ($\phi_{\rm Small}$) apertures are given for the SIS data.
See Section 5 for the details of the photon-collection regions.
No correction for vignetting and point spread function has been made.}
\begin{tabular}{ccc}
Detector & $\phi_{\rm Large}$ & $\phi_{\rm Small}$ \\
& $10^{-2}$\cps & $10^{-2}$\cps \\
S0 & 6.5 & 3.2 \\
S1 & 4.7 & 2.7 \\
G2 & 3.8 & --- \\
G3 & 6.1 & --- \\
\end{tabular}
\end{center}
\end{table}

NGC4395 was observed with ASCA on 1998 May 24--25 for a half day. 
The two Solid state Imaging Spectometers (SIS; S0 and S1) were operating in
1CCD Faint mode throughout the observation.
The best calibrated CCD chip on each detectetor (S0C1 and S1C3) was used.
The field in the vicinity of NGC4395 is remarkably
crowded with bright X-ray sources (e.g., see the ROSAT PSPC image by
Radecke 1997).  
The 1CCD mode observation restricted the SIS field of view to
a $11\times 11$ arcmin box which
covers the nucleus of NGC4395 and four other soft X-ray sources
detected with the ROSAT PSPC. 
The Gas Imaging Spectrometer (GIS; G2 and G3) 
has a larger field of view ($\sim 40$ arcmin in diameter) in which
at least four more X-ray sources are significantly detected.
These sources all have soft X-ray counterparts detected with the PSPC
(Radecke 1997).

The data reduction was carried out using FTOOLS version 4.2
and standard calibration
provided by the ASCA Guest Observer Facility (GOF) at Goddard Space 
Flight Center.
The pointing error of the ASCA satellite induced by the distortion of
the base plate of the star tracker has been corrected so that the pointing
accuracy in the ASCA images presented in this paper is the order of 10 arcsec.
The good exposure time is about 21 ks for each detector.
The mean count rates of NGC4395 obtained from 
the four detectors are summarised in Table 1.
Response matrices for the SIS were generated 
by SISRMG version 1.1. Version 4.0 of the redistribution matrices
provided by the GIS team are used for the GIS.
The effective areas of the source spectra were computed with
ASCAARF version 2.73.


\begin{figure}
\centerline{\psfig{figure=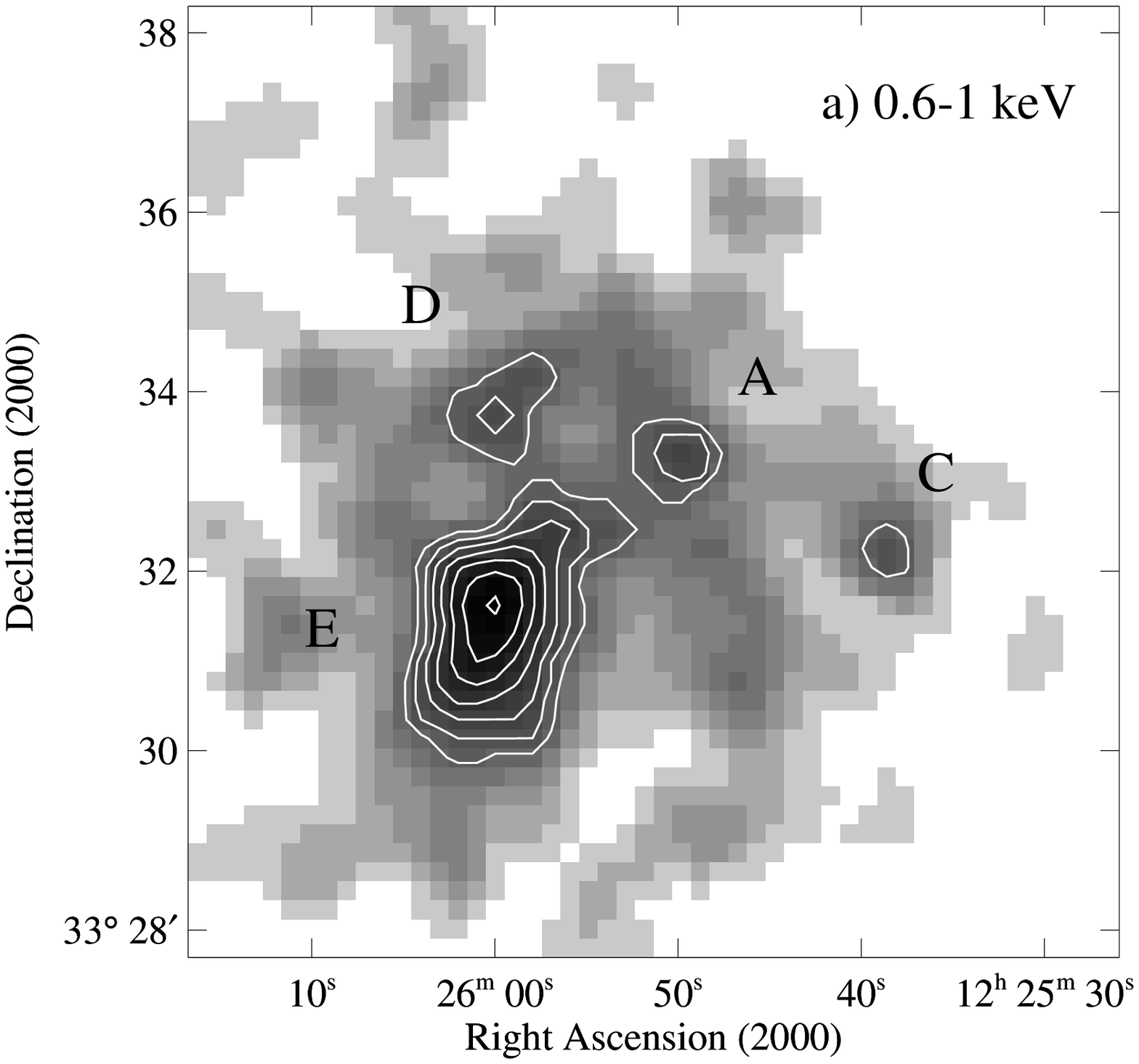,width=0.4\textwidth,angle=0}}
\centerline{\psfig{figure=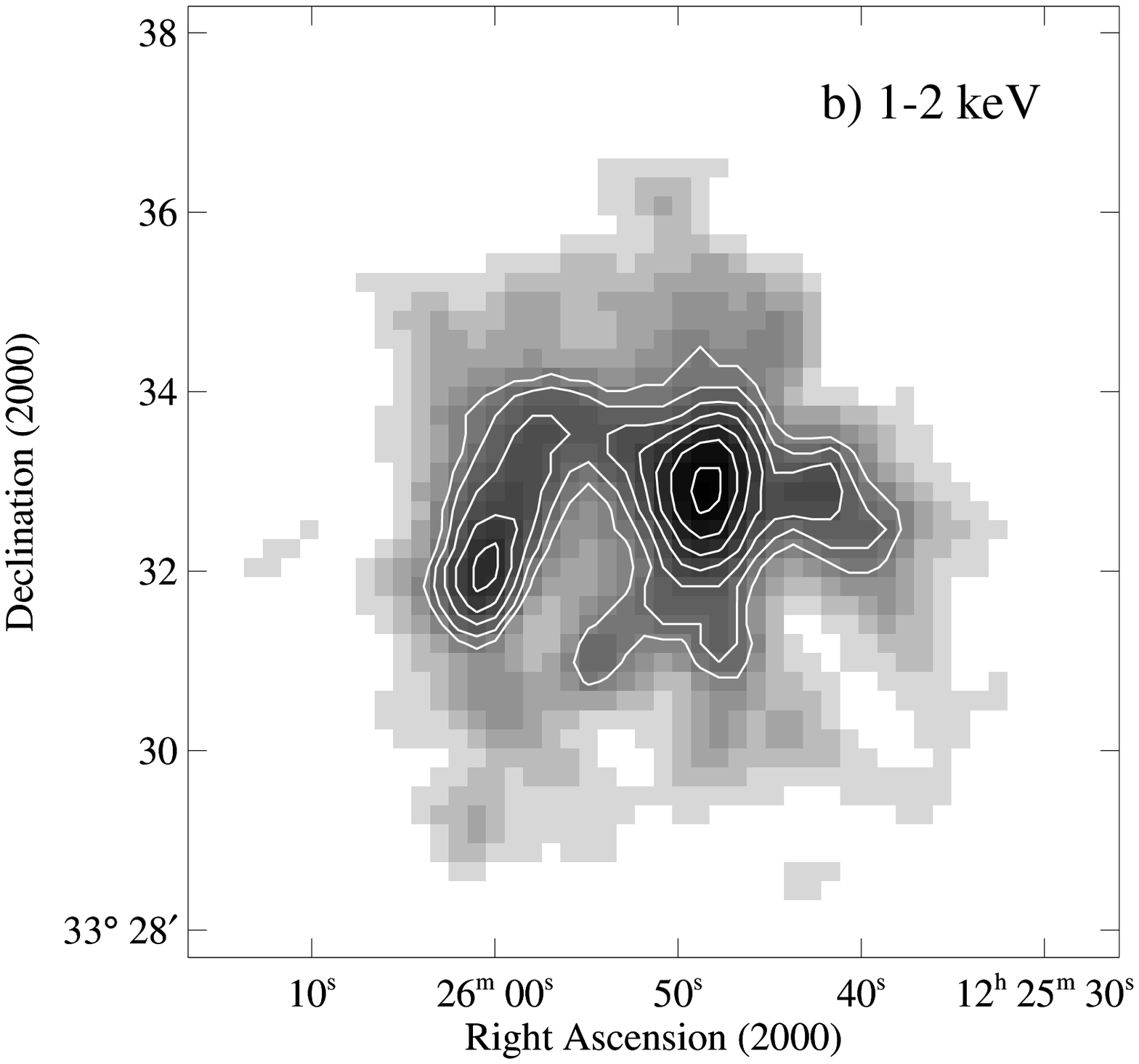,width=0.4\textwidth,angle=0}}
\centerline{\psfig{figure=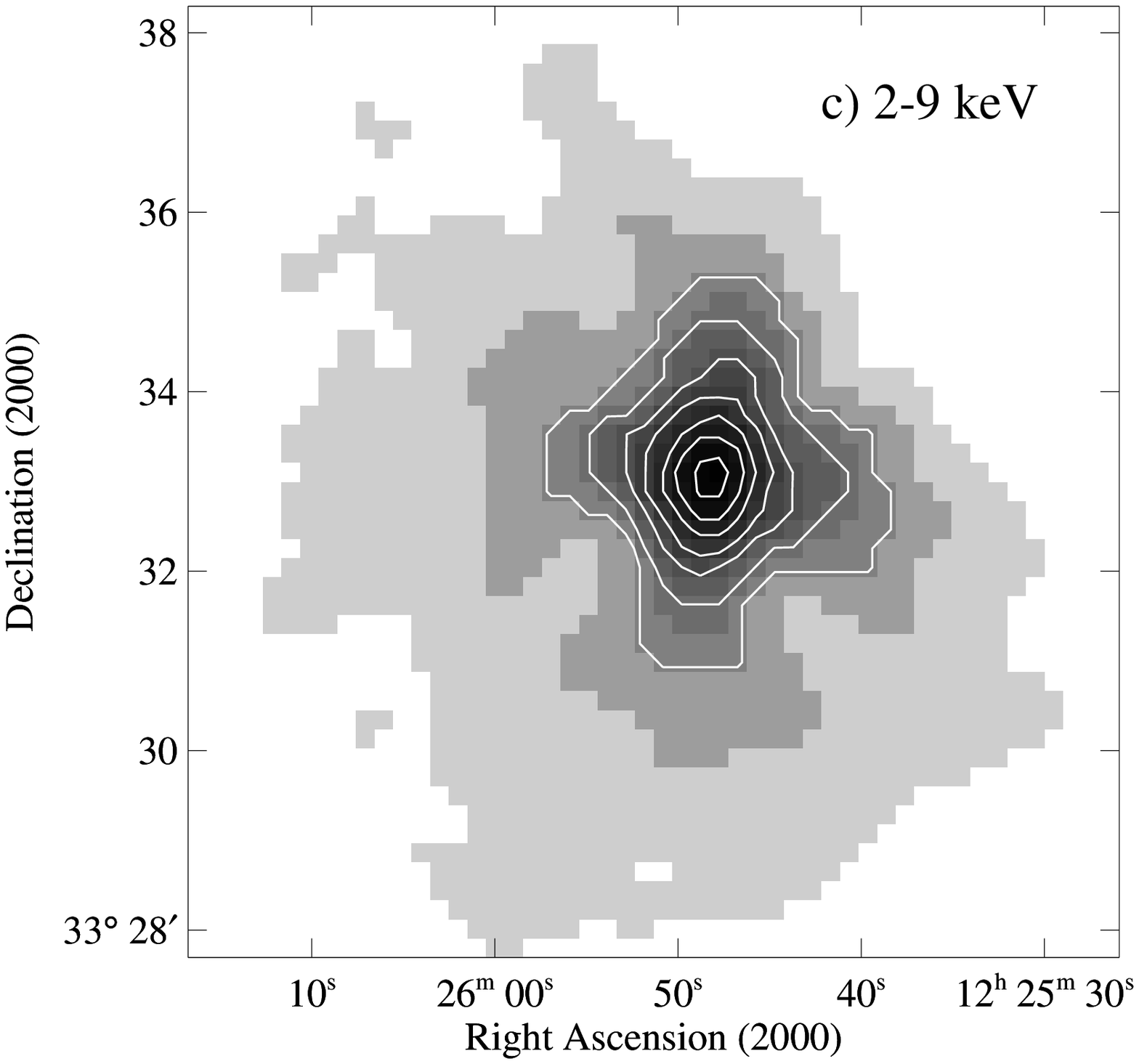,width=0.4\textwidth,angle=0}}
\caption{The ASCA SIS images of the NGC4395 region in the energy bands
of a) 0.6--1 keV, b) 1--2 keV and c) 2--9 keV. Seven contour levels
are drawn, equally spaced in intensity, between the peak and $2\sigma $ level of background. The
ROSAT PSPC sources, A, C, D and E (see Fig. 1 in Moran et al 1999 for
naming convention) are seen in the 0.6--1 keV band. The nuclear source
A becomes prominent in the 1--2 keV band and it is the only
significant source in the 2--9 keV band.}
\end{figure}

\section{X-ray images}

Five sources have been detected within 3 arcmin from the nucleus of
NGC4395 in the ROSAT PSPC image (Moran et al 1999), and we use the
same naming convention (A, B, C, D and E) for the five X-ray sources
as used by Moran et al (1999, see Fig. 1 in their paper). Since the
separations between the sources are not sufficiently large compared to
the Point Spread Function (PSF) of the ASCA X-ray Telescope (XRT with
the half-power diameter of 3 arcmin, Serlemitsos et al 1995), these
sources are resolved more clearly in the ROSAT PSPC image of higher
spatial resolution.  The position of source A coincides with the
optical nucleus of NGC4395. The brightest source E is found to have a
distinctly softer spectrum than the other sources.

The three ASCA SIS images in the energy bands of 0.6--1 keV, 1--2 keV
and 2--9 keV are shown in Fig. 1.  The data from the two SIS detectors
have been added together.  In the 0.6--1 keV band image, all the PSPC
sources except for B are seen (B is too faint or unresolved with
ASCA). E is the brightest source in the low energy band, as in the
PSPC image. However, the nuclear source A becomes brighter than E in
the 1--2 keV band. C and D are also visible.  Finally, in the 2--9 keV
band, A is the only source detected in the field.  This indicates that
the nuclear source of NGC4395 is bright above the ROSAT band and thus
has a hard spectum.  The brightest ROSAT source, E, declines steeply
around 2 keV in intensity and emits little X-ray emission above that
energy.

\section{X-ray variability}

\subsection{Light curve and timing analysis}


\begin{figure*}
\centerline{\psfig{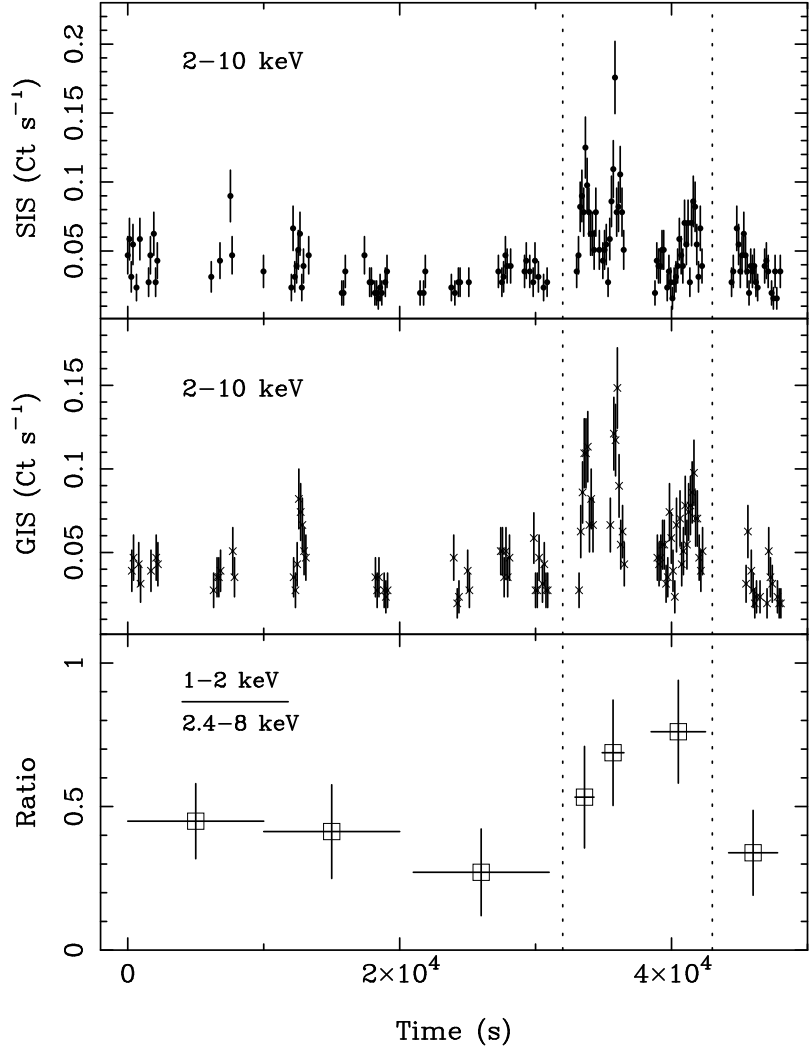}}
\caption{Top and middle panels: the 2--10 keV band light curves of
source A in NGC4395 obtained from the ASCA SIS and GIS; bottom panel:
the SIS count rate ratio in the 1--2 keV to 2.4--8 keV bands obtained
from the small aperture.  The epoch of the light curves is 1998 May
24, 16h32m25s (UT).  The active phase ($3.2\times 10^4$--$4.3\times
10^4$ s), during which the X-ray source exhibits flares, is indicated
in the figure.  The mean (1--2 keV)/(2.4--8 keV) count-rate ratios
during the active and quiescent phases are $0.66\pm 0.10$ and $0.37\pm
0.07$, respectively, showing significant spectral softening during the
active phase (see also Fig. 4 and Fig. 5), although a $\chi^2$ test
for the data points plotted in the bottom panel does not rule out a
constant hypothesis.}
\end{figure*}


The source photons of the SIS are collected from a 
$6\times 6$ arcmin box centred on source A, excluding a circular
region with a radius of 1 arcmin centred on E which is $\sim 2.8$ arcmin
away from A. 
This photon-collection region still contains B, C and D.
Since the exclusion of the central part of the E image 
discards only 40 per cent of 
the total photons from E as a result of the broad PSF 
of the ASCA XRT (Serlemitsos et al 1995), 
about 30 per cent of 
the photons from E should also spread over the region.
We attribute 
as much as 70 per cent of the observed SIS counts in the 0.6--1 keV band
to contamination by examining the ROSAT image 
in conjunction with the PSF of the ASCA XRT.   
Since the image analysis shows that the nuclear source of NGC4395 
dominates the energy band above 2 keV, the data collected from this 
region is presumed to be free from contamination in the energy band
above 2 keV.

Although the two bright sources, A and E, are resolved in the GIS
image, cross contamination between the two sources is severer than for
the SIS due to the detector's poorer spatial resolution. Therefore we
used a simple circular region with a radius of 5 arcmin centred on
source A, discarding the data below 2 keV.


The 2--10 keV SIS and GIS light curves with 128 s bins
are shown in Fig. 2. Those light curves were produced in
the same procedure described in Nandra et al (1997) and 
the two detectors of the same type of instruments are co-added.
X-ray flux changes are seen on various time scales from a hundred
seconds to half-day of the whole observing run.
The shortest doubling time is only $\sim 100$ s. 
At least three flares with a duration of $\sim 1000$ s are seen 
in a time interval, $3.2\times 10^4$--$4.3\times 10^4$ s, 
indicated in Fig. 2 (hereafter this time interval is called ``active''
state and the rest is ``quiescent'' state).

We have verified that the position of the brightest X-ray source
during the ``active'' period coincides with the position of source A,
or the optical nucleus of NGC4395. This rules out the possibility that one of
the nearby X-ray sources is responsible for the X-ray flaring.


The normalized excess variance, $\sigma^2_{\rm RMS}$, 
was defined in Nandra et al (1997)
as a measure of variability amplitude.
(Note that Turner et al (1999) have pointed out an mistake in the
formula quoted by Nandra et al (1997) for the error on $\sigma^2_{\rm RMS}$.
We have used the corrected formula given by Turner et al 1999.)
We computed $\sigma^2_{\rm RMS}$ for the 2--10 keV light curves and 
found $0.203\pm 0.066$ for the SIS and $0.176\pm 0.047$ for the GIS.


\begin{figure}
\centerline{\psfig{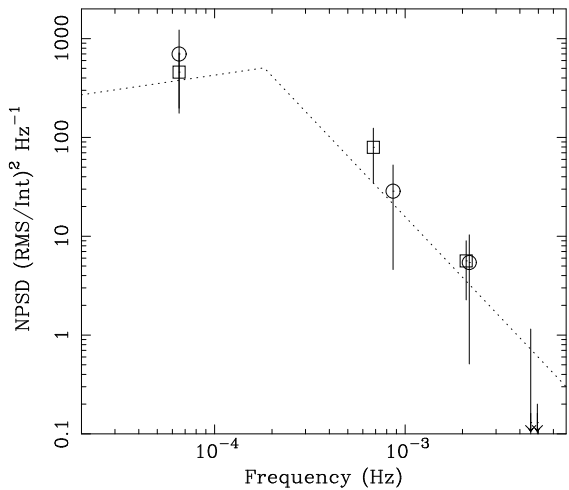}}
\caption{The normalized power spectrum density (NPSD) obtained from
the SIS (open squares) and GIS (open circles) light curves. See
Hayashida et al (1998) for the definition of the NPSD and the details
of the computing method. Poisson noise has been subtracted. Negative 
values are found for both detectors at the 
highest frequencies ($-1.03\pm 2.18$ 
at $4.56\times 10^{-3}$ Hz for the SIS and $-2.13\pm 1.87$ at 
$4.94\times 10^{-3}$ Hz for the GIS) where the Poisson noise 
overwhelms the signal. The dotted
line is the broken power-law fitted to the data, using the shape of 
the NPSD for NGC4051 (see Section 6.3 for details).}
\end{figure}

Our light curve data are not sufficient to calculate a fully sampled
power spectrum by standard means. However we estimate the spectral
properties by two methods. First, we estimate the power spectrum using
the algorithm of Lomb (1976) which is specially designed for
irregularly sampled data. 
The result shows the red-noise character typical of more
luminous AGN. 
Next, we make a
quantitative comparison to other AGN, by calculating the normalized
power spectral density (NPSD), following the methodology of Hayashida
et al (1998) at a small number of frequencies, which is presented in Fig. 3. The error
bars on each data point are too large to constrain the slope of the
power spectrum. However the amplitude of variability in NGC4395 is
comparable to that of NGC4051 observed with Ginga (Hayashida et al
1998).

\subsection{Spectral variability}


\begin{figure}
\centerline{\psfig{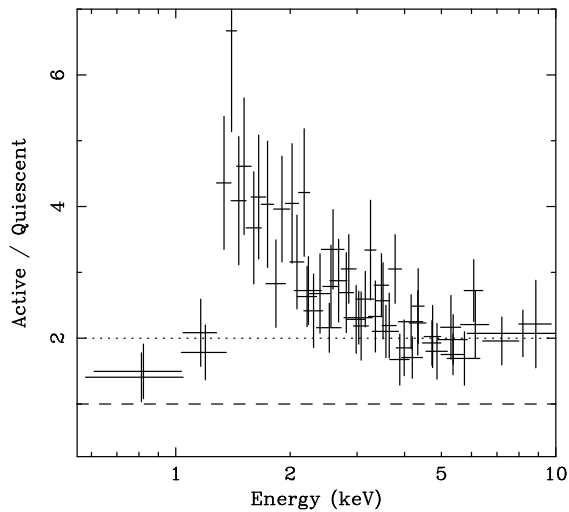}}
\caption{Spectral ratio of the data from the 
active state to the quiescent state.
The small aperture SIS data and the GIS data above 2 keV are plotted.
The dashed line indicates a ratio of unity while the dotted line 
shows a factor of two change which is consistent above 3 keV.}
\end{figure}


\begin{figure}
\centerline{\psfig{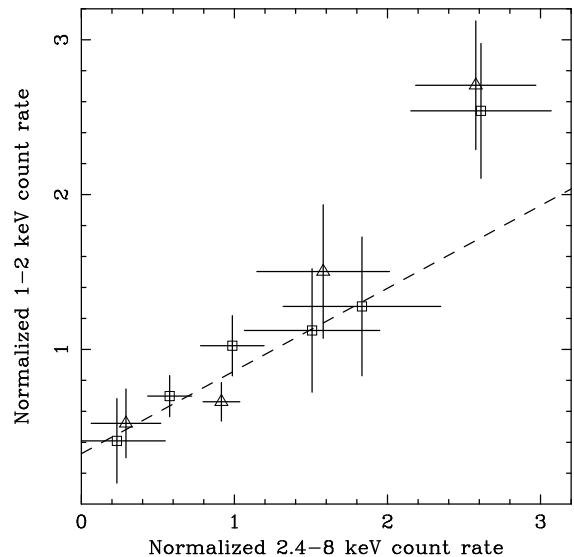}}
\caption{Plot of normalized count rates in the 1--2 keV band against 
those in the 2.4--8 keV band obtained from the S0 and S1 detectors 
for the six (S0; squares) and four (S1; triangles) 
count-rate slices, respectively, taken from each 2--10 keV light curve with 
128~s bins. The dashed line shows the best-fit correlation for 
the data points, excluding the highest flux bins.}
\end{figure}


In order to investigate the soft X-ray spectrum (below 2 keV), 
data restricted within a small aperture around the nuclear source
are taken from the SIS detectors so that the contamination is
minimized.  
A circular region with a 3 arcmin diameter, which corresponds
to the half-power diameter of the ASCA XRT PSF, is used instead
of the original photon collection region.
A contribution from the nearby sources in the band below 1 keV is 
expected to be about 10 per cent, but should be negligible in the 1--2
keV band. 

The count rate ratio in the 1--2 keV to 2.4--8 keV bands obtained
from the small aperture SIS data is plotted
in Fig. 2. 
The excess variation in the soft band relative to the hard band 
is confirmed by the spectral
ratio between the active and quiescent states (Fig. 4).
The data are consistent with a flux change by a factor of 2 with
no spectral variation between the two states above 3 keV 
while a large excess variation is evident in the 1--2 keV band.

The X-ray source shows a rapid, large amplitude flux variations 
even within the active state. 
We next investigated spectral variability associated with the rapid
flux variations, using a normalized
count rate diagram (Fig. 5), similar to that used 
in Papadakis \& Lawrence (1995) for NGC4051.

The whole 2--10 keV light curves of the S0 and S1 detectors were 
divided into six and four 
count-rate slices, respectively. Count rates from each slice in 
the 1--2 keV and 2.4--8 keV bands were then normalized by the 
mean count rates in the respective energy band.
This plot could, in principle, track 
spectral variability on a time scale down to the time resolution of the
light curve, 128 s.
As Fig. 5 shows, a significant excess variation in the 1--2 keV band
occurs at the highest flux bin. 
Since this bin contains all the three flare-peaks in the active state,
the spectral softening appears to be associated with the flaring 
and may track it with a time lag of less than $10^3$ s.

\section{X-ray spectrum}

As the image analysis shows, the X-ray spectrum of the nucleus of
NGC4395 is hard, probably due to absorption.  We first investigate the
2--10 keV data to determine the power-law continuum. Since the
contamination from the nearby sources is negligible in this energy
band, the large aperture data from both SIS and GIS integrated over
the whole observation are used (Section 5.1).

On investigating the lower energy part of the spectrum, we use only
the small aperture data from the SIS to avoid the contamination. A
simple absorbed power-law model fails to explain the soft X-ray data
due to the presence of excess emission. The soft excess emission is
associated with the NGC4395 nucleus itself and possible origins are
discussed (Section 5.2).

The low energy band spectrum shows a dramatic change during the active
state (see Section 4.2). We examine the spectra during the quiescent
and active states and try to explain the spectral change between the
two states with a variable warm absorber model (Section 5.3).

The spectral analysis presented here was performed using 
XSPEC version 10.0. Quoted errors on spectral parameters are 
90 per cent confidence region for one parameter of interest,
unless stated otherwise.

\subsection{Time-averaged 2--10 keV spectrum}


\begin{figure}
\centerline{\psfig{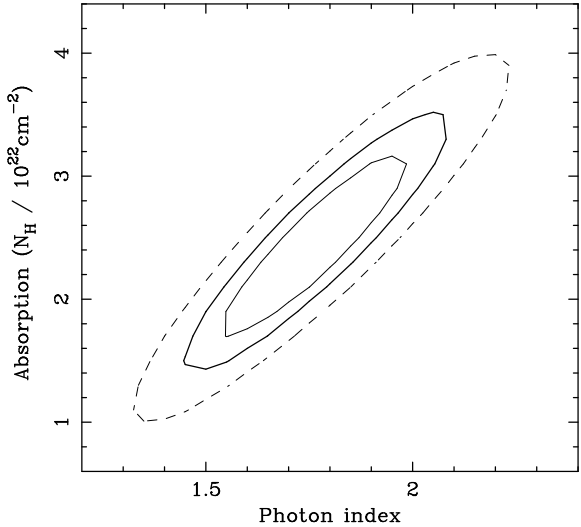}}
\caption{Confidence contours in the plane of photon-index against absorption
column density obtained from a power-law fit to the total, time-averaged
2--10 keV data 
from the four detectors. The absorption is assumed to occur in 
neutral gas. The contours are drawn at the 68, 90 and 99 per cent
confidence levels for two parameters of interest.}
\end{figure}


\begin{figure}
\centerline{\psfig{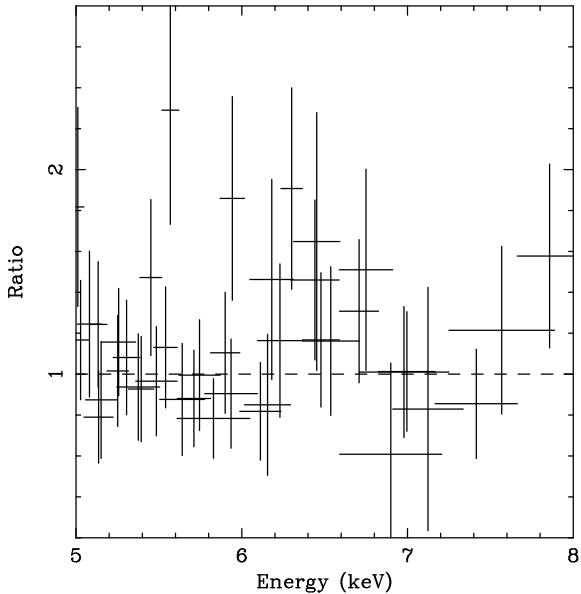}}
\caption{The ASCA SIS and GIS data of NGC4395 at the Fe K band divided by 
the best-fit power-law model for the neighbouring continuum. 
A Fe K line is detected at 6.45 keV marginally above
the 90 per cent confidence level.}
\end{figure}

Fitting jointly the 2--10 keV data from the SIS and GIS 
with an absorbed power-law gives a photon-index $\Gamma =
1.72^{+0.24}_{-0.27}$ and column density \nH\ $=2.3^{+0.8}_{-0.9}
\times 10^{22}$\psqcm\ (see confidence contours in Fig. 6).
Here the absorption is assumed to be neutral and the absorption
cross-sections taken from Morrison \& McCammon (1983) are used.
The higher energy continuum may be slightly steeper, e.g., $\Gamma =
2.1\pm 0.3$ for the 4.5--10 keV data when \nH\ is fixed
at $2.3\times 10^{22}$\psqcm.
A narrow iron K line is marginally detected at $6.45^{+0.28}_{-0.23}$ keV with 
an equivalent width (EW) of $180\pm 150$ eV (Fig. 7).
The observed 2--10 keV flux is $4.5\times 10^{-12}$\ergpspsqcm,
and the absorption-corrected 2--10 keV luminosity is $4.1\times 10^{39}
(D/2.6{\rm Mpc})^2$\ergps, where $D$ is the distance to the
galaxy.

\subsection{Variable soft X-ray spectrum and its possible origins}


\begin{figure}
\centerline{\psfig{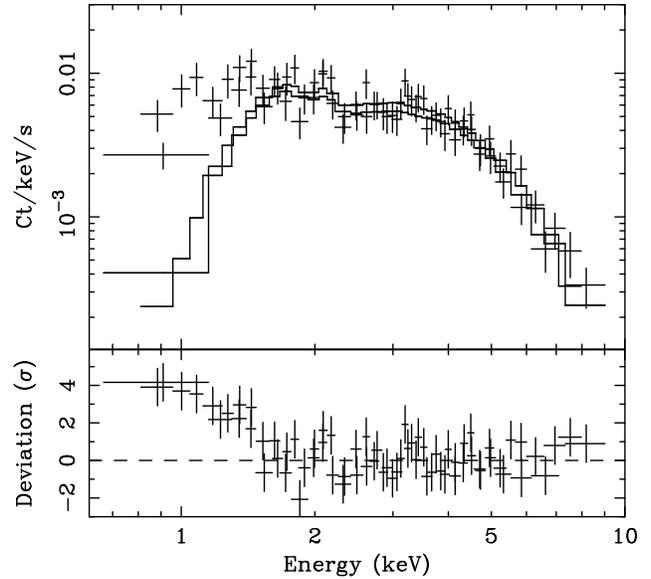}}
\caption{Upper panel: the time-averaged small-aperture spectrum of 
NGC4395 obtained from the ASCA SIS;
lower panel: residuals in $\sigma$ when the data are compared with
the absorbed power-law model best-fitting the large-aperture 2--10 keV data 
($\Gamma=  1.72$, \nH $=2.3\times 10^{22}$\psqcm) indicated in
solid histograms in the upper panel. Surplus emission in the soft X-ray
band is evident.}
\end{figure}

An extrapolation of the model best-fitting the 2--10 keV data 
to the lower energy leaves 
significant surplus emission below 2 keV (in the small aperture data, Fig. 8).

Here we discuss several possibilities for 
the origin of the excess soft X-ray emission.
Thermal emission or reflected light of the obscured active nucleus
from an extended region are readily ruled out by the rapid 
variability observed in the 1--2 keV band.

It could be a soft excess component intrinsic to the primary source 
like the one seen in the X-ray spectrum of Seyfert 1 galaxies and quasars
and obscured by the same matter which attenuates the primary power-law.
Assuming the soft excess has a blackbody type spectral form, 
a fit to the small-aperture spectrum together with the power-law
sharing the same absorption (\nH $\sim 2\times 10^{22}$\psqcm) 
suggests the blackbody to have a temperature $kT =0.08\pm 0.02$ keV.
However, such obscured blackbody emission is argued against by the 
the energetics. 
The soft X-ray luminosity of the blackbody
component would be $3\times 10^{42}$\ergps, when corrected for absorption,
and 3 orders of magnitude above the power-law component.
It would produce far more luminous optical/UV emission-lines
and far-infrared dust reradiation emission than observed.

Reflection from a partially ionized accretion disk could emerge in this
energy range. However, the soft X-ray variation which is larger 
than that of the primary continuum 
is hard to explain by this hypothesis.

A possible solution is to introduce a warm absorber which varies in
response to the continuum source.
This model needs no extra X-ray source but does need 
different physical conditions for the X-ray absorber.
Unlike many higher luminosity Seyfert 1 galaxies in which OVII (0.74 keV)
and OVII (0.85 keV) are major features of the warm absorption, the primary
absorption occurs around 2 keV in NGC4395.
It requires a higher ionization parameter and the absorption is due to
highly ionized O, Ne, Mg, Si, S and Fe-L.

Introducing a warm absorber reduces the opacity of the cold absorption
significantly.
This is more consistent with the observed optical/UV properties, e.g.,
non-stellar UV continuum, broad permitted line emission, high 
excitation lines usually seen in Seyfert 1 nuclei.
The column density implied from cold absorption alone fitted to 
the 2--10 keV spectrum
corresponds to $A_{\rm V}\sim 10$ when the Galactic gas-to-dust ratio is used.
The Seyfert-1 like optical/UV properties would not be observed 
if the central source and the broad-line region (BLR) are behind such 
a heavy obscuration, although some reddening to the BLR and the central source
is still required given the ionizing photon deficit (e.g., for narrow 
H$\beta$), 
pointed out by Lira et al (1999), Kraemer et al (1999) and 
Moran et al (1999),
the Seyfert 1.8/1.9 nature in the optical emission-line spectrum (e.g.,
Ho et al 1997) and the large H$\alpha$/H$\beta$ ratio 
($\sim 5.1$, Kraemer et al 1999) for the broad component.

As the soft X-ray part of the spectrum has such a variable nature,
a spectral variability study is more useful in modelling the soft X-ray
spectrum than using the time-averaged spectrum alone. 
The warm absorber hypothesis is then explored to account for the 
spectral variability between the active and quiescent states
in the next section.

\subsection{Variable warm absorber}

We try to model the spectral change in the soft X-ray band between
the quiescent and active states with a warm absorber. The continuum
source is assumed to have a constant spectral shape of a power-law with
photon-index $\Gamma = 1.72$, obtained from the integrated 2--10 keV
data, but different luminosities in the two states.
Since it is unlikely that column density 
changes in response to the continuum source, the ionization parameter,
$\xi=L/(nR^2)$ of absorbing matter is a primary driver of the spectral
change. We used the warm absorber model {\tt absori} by Done et al
(1992) in XSPEC. 
Cold absorption is also included in the model.
The small aperture SIS spectra of the quiescent and active states are
fitted jointly to obtain parameters which are shared between the two datasets,
such as a column density of an absorber.
The mean absorption-corrected 2--10 keV fluxes of the power-law continuum
during the quiescent and active states are 
$3.2\times 10^{-12}$\ergpspsqcm\ and 
$6.1\times 10^{-12}$\ergpspsqcm, respectively. 

\subsubsection{Single warm absorber model} 

A model of a single warm absorber with variable $\xi$ 
plus a constant cold absorber is first fitted.
The temperature of the absorbing gas is assumed to be $3\times 10^5$ K.
Column densities of warm and cold absorbers (\nW\ and \nH, respectively)
are free parameters but set to vary in unison between the two datasets.
The ionization parameter ($\xi$) of the absorber is allowed to take
different values in the two datasets as well as the normalization of 
power-law. This parameter setting enables the model to accommodate 
a physically reasonable change in parameters in a variable warm absorber
hypothesis.
This model, however, gives a poor fit to the absorption band (below
2 keV). Despite the flux change of more than a factor of two, the ionization
parameters obtained from the fit differ very little (Model-1 in Table 2). 

When the values of $N_{\rm W}$ for the two datasets are also 
allowed to vary independently, the fit improved significantly
(Model-2 in Table 2).
However, not only has $N_{\rm W}$ decreased in the active state, but 
the lower $\xi$ in the active state is in the opposite sense to 
the photoionization hypothesis.
Therefore the single warm absorber model fails, but the latter fit
instead suggests the existence of two physically distinct absorbers, i.e,
the absorption feature in the two spectra are dominated by different absorbers.
Next we try a multi-layer warm absorber model.


\begin{table*}
\begin{center}
\caption{Fits of the single warm absorber model to the quiescent and 
active state spectra. The photon-index of power-law, $\Gamma = 1.72$,
is assumed for the two datasets. The warm absorber model {\tt absori}
(Done et al 1992) is used. The temperature of the
absorber is assumed to be $3\times 10^5$ K.
{\it Q} and {\it A} denote the values for the quiescent and active state
data, respectively. In Model (1), values for \nW\ are tied together between
the two datasets while in Model (2), \nW are allowed to vary independently.}
\begin{tabular}{cccccc}
Model & $T$ & \nW & $\xi$ & \nH & $\chi^2$/dof \\
& K & $10^{22}$\psqcm & \ergcmps & $10^{21}$\psqcm & \\[5pt]
(1) & $3\times 10^5$ & $7.0^{+1.7}_{-1.7}$ & {\it Q}: $92^{+17}_{-18}$/{\it A}: $102^{+24}_{-22}$ & $5.1^{+1.5}_{-1.1}$ & 75.63/67 \\
(2) & $3\times 10^5$ & {\it Q}: $10.9^{+2.7}_{-2.3}$/{\it A}: 
$2.5^{+1.1}_{-0.9}$
& {\it Q}: $109^{+18}_{-18}$/{\it A}: $21^{+25}_{-15}$ & $3.3^{+1.3}_{-0.9}$ 
& 50.42/66 \\
\end{tabular}
\end{center}
\end{table*}


\begin{table*}
\begin{center}
\caption{Multi-zone warm absorber fit to the quiescent and 
active state spectra. One absorber is assumed to be constant in 
both states while the other is variable only in ionization parameter
($\xi_{\rm v}$).
All the parameters of the two warm absorbers apart from $\xi_{\rm v}$ 
are shared between the two datasets.
The photon-index of power-law, $\Gamma = 1.72$,
is assumed. The temperatures of the
absorbers are assumed to be $1\times 10^5$ K for the constant absorber
and $1\times 10^6$ K for the variable one.
{\it Q} and {\it A} denote the values for the quiescent and active state
data, respectively.}
\begin{tabular}{cccccccc}
\multicolumn{3}{c}{Constant} & \multicolumn{3}{c}{Variable} & Cold & \\
$T_{\rm c}$ & $N_{\rm W}^{\rm c}$ & $\xi_{\rm c}$ &
$T_{\rm v}$ & $N_{\rm W}^{\rm v}$ & $\xi_{\rm v}$ &
\nH & $\chi^2$/dof \\
K & $10^{22}$\psqcm & \ergcmps & 
K & $10^{22}$\psqcm & \ergcmps & 
$10^{21}$\psqcm & \\[5pt]
$1\times 10^5$ & $2.4^{+1.0}_{-0.7}$ & $31^{+36}_{-24}$ &
$1\times 10^6$ & $9.9^{+5.0}_{-4.0}$ &
{\it Q}: $140^{+90}_{-45}$/{\it A}: $\geq 480$ & $2.9^{+2.1}_{-2.8}$
& 55.75/65 \\
\end{tabular}
\end{center}
\end{table*}

\subsubsection{Multi-zone warm absorber model}


\begin{figure}
\centerline{\psfig{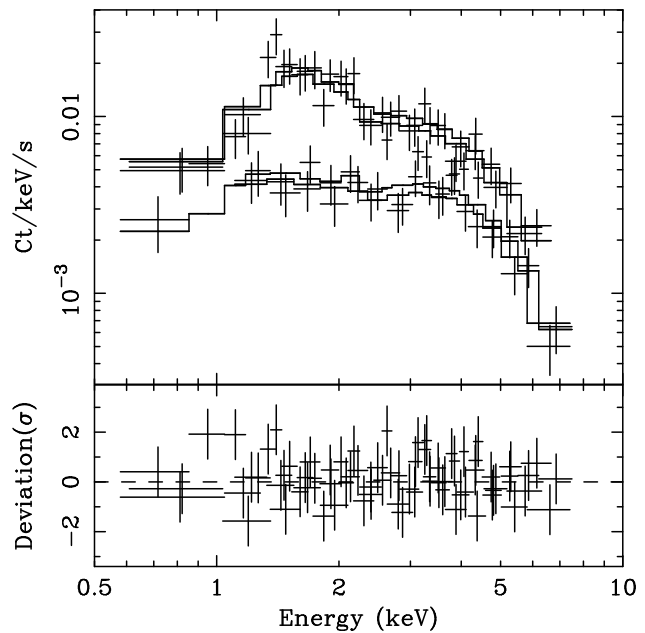}}
\caption{The ASCA SIS spectra of NGC4395 in the active (above)
and quiescent (below) states fitted jointly 
with the multi-zone warm absorber model (Table 3). The contour levels are 
68 and 90 per cent confidence regions for two parameters of interest.
The ionization parameter of the 
variable absorber changes between the active and quiescent datasets
in response to the continuum luminosity variation 
while the column density is assumed to remain the same.
The constant absorber is common to
the two datasets.}
\end{figure}


\begin{figure}
\centerline{\psfig{figure=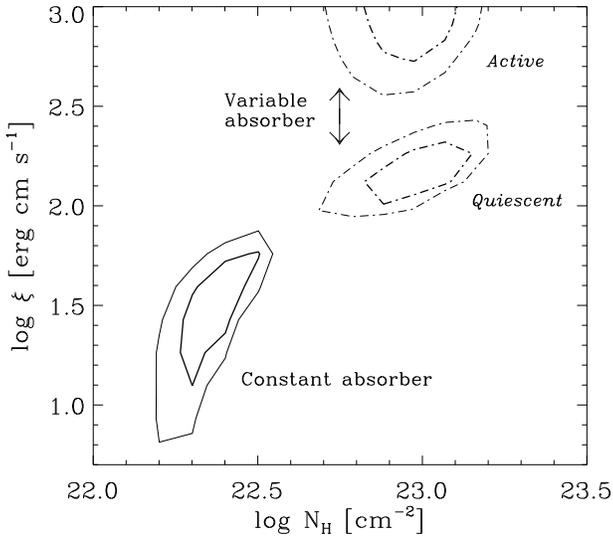,width=0.57\textwidth,angle=0}}
\caption{Confidence contours in the plane of column density (\nW) and 
ionization parameter ($\xi$) of the constant and variable warm absorbers
when the multi-zone absorber model is fitted to the active and 
quiescent datasets jointly (see Table 4). The contour levels are 
68 and 90 per cent confidence regions for two parameters of interest.
The ionization parameter of the 
variable absorber changes between the active and quiescent datasets
in response to the continuum luminosity variation 
while the column density is assumed to remain the same.
The constant absorber is common to
the two datasets.}
\end{figure}

In addition to a variable warm absorber, a constant
warm absorber, which is primarily seen in the active state, has been
introduced. All the parameters of the two warm-absorbers are shared by
the two datasets apart from the ionization parameter of the variable 
absorber ($\xi_{\rm v}$). 
The temperatures of the constant and variable absorbers are 
assuemd to be $1\times 10^5$ K and $1\times 10^6$ K, respectively.
This model gives a good fit to the data despite the only one parameter
($\xi_{\rm v}$) in the two warm-absorbers differing between the two datasets
(Table 3, Fig. 9, Fig. 10). 

The ionization parameter ($\xi_{\rm c}\simeq 30$ \ergcmps) and column
density ($N_{\rm W}^{\rm c}\simeq 2\times 10^{22}$\psqcm) of the
constant absorber are found to be similar to those seen in higher
luminosity Seyfert 1 nuclei (Reynolds 1997; George et al 1998).  The
variable absorber has a higher and variable ionization parameter
($\xi_{\rm v}$) and a larger column density ($N_{\rm W}^{\rm v}\simeq
1\times 10^{23}$\psqcm).  It imposes an absorption feature above 1 keV
and peaked around 1.5 keV in the quiescent state while it has almost
disappeared in the active phase due to high ionization ($\xi_{\rm
v}\geq 480$ \ergcmps).  The change in $\xi_{\rm v}$ is marginally
consistent with the change in luminosity of the continuum source which
has shown a factor of 2--3 variations during the active phase.  The
multi-layer warm absorber model makes excess cold absorption
marginally required at the 90 per cent significant level (\nH $ =
2.8^{+2.1}_{-2.8}\times 10^{21}$\psqcm).  The 0.5--2 keV fluxes for
the quiescent and active states estimatd from this model are
$2.0\times 10^{-13}$\ergpspsqcm\ and $6.2\times 10^{-13}$\ergpspsqcm,
respectively.  If the covering factor of the absorbing gas is high,
emissin-lines from the ionized gas in the absorber are expected.
There is an emission-line like feature at 1.4 keV seen in the active
state spectrum, which can be identified with MgXI.

\section{discussion}

\subsection{Complex absorption in the nucleus of NGC4395}

The 2--10 keV ASCA spectrum is well described by a power-law 
of $\Gamma\simeq 1.72$ modified by cold absorption of \nH $\simeq 2.3\times
10^{22}$\psqcm. This cold absorption model however does not explain
the soft X-ray spectrum (see Fig. 8). 
A variability study revealed that emission in the 1--2 keV band  
is more variable than in the higher energy band (Fig. 4).
The variation above 2 keV is directly attributed to the intrinsic 
flux change in the primary source.
A plausible explanation for the excess variability in the 1--2 keV band
is a change in a warm absorber. 

Although Moran et al (1999) speculated about the presence of a warm absorber
through spectral analysis of the ROSAT PSPC data, the properties of the
X-ray absorption in NGC4395 appear to be more complex than they
assumed. As shown by the ASCA spectrum, the ROSAT energy range 
is dominated by absorption, which makes the use of 
the PSPC spectrum to assess the primary continuum difficult.

A spectral fit to the ASCA data from the quiescent and active states
can be explained by the presence of a constant and a variable warm absorber.
The physical condition of the constant absorber is similar to
that seen in many higher luminosity Seyfert 1 galaxies (Reynolds 1997;
George et al 1998) while the 
variable absorber is found to have higher ionization parameter, which
leads absorption features to be imprinted around 2 keV due to highly
ionized O, Ne, Mg, Si, S and Fe-L. 
Evidence for multi-zone warm absorber has also been found in Seyfert 1
nuclei like MCG--6-30-15 (Otani et al 1996; Morales, Fabian \& Reynolds 1999)

Using the formulae in Otani et al (1996) for the warm absorber in 
MCG--6-30-15, the receombination time scale for highly ionized gas
can be approximated by
$t_{\rm rec}\simeq 200n_9^{-1}T_5^{0.7}(Z/Z_{\rm O})^{-1}$ s, 
or $t_{\rm rec}\simeq 200\xi_2L_{40}^{-1}T_5^{0.7}R_{-4}^2(Z/Z_{\rm O})^{-1}$ 
s, where density is $10^{9}n_{9}$ \pcubcm,
temperature is $10^5T_5$ K, ionization parameter is $100\xi_2$ erg\thinspace
cm\thinspace s$^{-1}$, the luminosity of the irradiating source is
$10^{40}L_{\rm 40}$ \ergps, the distance of the warm absorber from the 
irradiating source is $10^{-4}R_{-4}$ pc, and $Z$ is the atomic number
($Z_{\rm O} = 8$ for oxygen). 
The luminosity of the continuum source during the active state is
about $1\times 10^{40}$\ergps\ at the source distance of 2.6 Mpc, 
when the power-law is integrated over 13.6 eV to 20 keV.

If the constant absorber, which is dominated by 
oxygen absorption, is indeed constant 
during the active state for $10^4$ s, the recombination time scale is
then $t^{\rm c}_{\rm rec}\geq 10^4$ s.
This locates the absorber at the
distance $R_{\rm c}\geq 1.3\times 10^{-3}L_{40}^{0.5}T_5^{-0.35}$pc 
from the central source and constrains the density to be
$n_{\rm c}\leq 2\times 10^7T_5^{0.7}$\pcubcm.
On comparing with the model for the nuclear emission-line region 
derived from the photoionization calculation based on the optical/UV 
properties by Kraemer et al (1999), $R_{\rm c}$
is just outside the BLR (note that the size of the emission-line regions 
derived by Kraemer et al 1999 becomes larger if the ionizing luminosity
obtained from our work is used).
The absorption features seen across the CIV line profile presented 
in Kraemer et al (1999) could be due to this warm absorber, as the 
ionization parameter ($\xi\sim 30$ \ergcmps) is consistent with CIV.

The duration of the individual flares in the active state is about 1000 s.
Since the spectral softening in the absorption band appears to occur 
at the peaks of those flares (Fig. 6), the recombination time scale of 
the variable absorber is probably less than the duration of the individual
flares, i.e., $t_{\rm rec}^{\rm v}\leq 1000$ s.
This gives constraints on density and distance of the absorber:
$n_{\rm v}\geq 10^9T_6^{0.7}(Z/Z_{\rm O})^{-1}$\pcubcm, and 
$R_{\rm v}\leq 4.5\times 10^{-5}L_{40}^{0.5}T_6^{-0.35}(Z/Z_{\rm O})^{0.5}$
pc, when the temperature of the absorber is $T_{\rm v}=10^6T_6$ K.
As highly ionized Ne, Mg, Si and S as well as O are 
major elements for absorption, $(Z/Z_{\rm O})\simeq $ 1--2.
The filling factor is described as
$\Delta R/R= \xi N_{\rm W}R/L \simeq 3\times 10^{-2}\xi_2N_{\rm W23}R_{-5}L^{-1}_{40}$, 
where the distance of the absorber is $R = 10^{-5}R_{-5}$ pc and the 
column density is $N_{\rm W} = 1\times 10^{23}N_{\rm W23}$\psqcm.
When $\xi_{\rm v} = 500$ \ergcmps\ and $R_{\rm v} = 1\times 10^{-5}$ pc are 
assumed, the filling factor and the density are 
$\Delta R/R\simeq 0.15$ and $n_{\rm v}\simeq 2\times 10^{10}$\pcubcm.

If the absorbing matter is space filling as suggested by the above argument,
significant emission from the same matter is also expected. 
This is consistent with the detection of MgXI at 1.4 keV. 
Although the quality
of our data is not sufficient to investigate weaker line emission,
the emission could mask some absorption features which may cause
a supuriously high ionization parameter.


\subsection{Multi-wavelength energy distribution}


\begin{figure*}
\centerline{\psfig{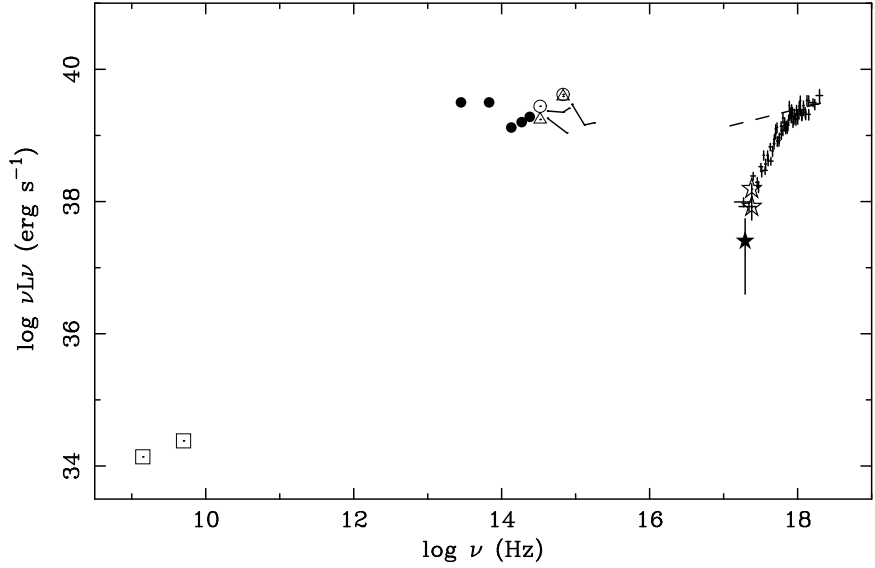}}
\caption{The spectral energy distribution of the nuclear source in 
NGC4395. The luminosities are calculated for a source distance of 2.6 Mpc.
The data are compiled from 
Lira et al 1999 (solid lines, open circles and open triangles: 
optical/UV data from WHT and HST, details therein; open stars: ROSAT PSPC;
filled star: ROSAT HRI) and Moran et al 1999 (open squared: VLA at 
6cm and 20cm; solid circles: ISO/SWS, see also Kraemer et al 1999, and
J, H, and K from Keck-I). The ASCA SIS 0.5--10 keV data are shown in crosses. 
The dashed line shows the unabsorbed
power-law of $\Gamma = 1.7$. This demonstrates that 
the intrinsic soft X-ray luminosity of the nuclear source 
is actually an order of magnitude larger than that estimated from the 
ROSAT observations (Kraemer et al 1999; Lira et al 1999; Moran et al 1999).}
\end{figure*}

It has been pointed out by Moran et al (1999, and also see Lira et al
1999) that the spectral energy distribution (SED) of NGC4395 diverses
from either a typical one of Seyfert/radio quiet quasars or of dwarf
AGNs. In particular, X-ray quietness measured with ROSAT has been
remarked. However, the absorption-corrected flux of the active nucleus
in NGC4395 at 1 keV obtained from the present work is $\simeq
2.2\times 10^{-12}$\ergpspsqcm, or $\simeq 1.7\times 10^{39}(D/2.6
{\rm Mpc})^2$\ergps\ in luminosity, an order of magnitude above the
previous ROSAT estimates, which came from a measurement of the
emission in the midst of the absorption band.  The SED of NGC4395 over
the radio to X-ray wave bands including our ASCA measurement is
plotted in Fig. 11. Corrected for the absorption, the active nucleus
in NGC4395 is no longer unusually X-ray quiet and the shape of the SED
may be more similar to that of Seyferts or radio-quiet quasars than
dwarf AGNs. The lack of an obvious ``big blue bump'' in the UV
spectrum is still a major difference from higher luminosity Seyferts
and quasars. This could be due to the small black hole mass inferred
for this galaxy ($\sim 10^5$\Ms, Kraemer e al 1999 and see the
following section), which would shift the thermal emission peak from
the accretion disk to shorter wavelengths thus it escapes being
observed.

The UV--soft X-ray luminosity ($10^{40}$-$10^{41}$\ergps), larger than
previously estimated when corrected for the obscuration implied from
the X-ray absorption, solves the problem of the ionizing photon
deficit for the optical narrow lines and thus supports the hypothesis
of a dust obscuration of the optical nucleus (Kraemer et al
1999). Part of the infrared emission detected by IRAS ($L_{\rm
FIR}\sim 1\times 10^{41}$\ergps) can be attributed to dust-reradiation
of the absorbed soft X-ray and UV photons, although the cold IRAS
colour ($S_{60}/S_{25}=11.6$, $S_{100}/S_{60}=2.7$) suggests that much
of the infrared emission is due to the galaxy disk (Lira et al 1999).

\subsection{X-ray variability and the black hole mass}

Strong X-ray flux variation was observed during the present ASCA
observation. This is consistent with an accreting black hole model
to account for the nuclear activity in NGC4395 (Filippenko, Ho
\& Sargent 1993) and 
strongly argues against other hypotheses like a starburst.
The large amplitude flux changes on a short 
time scale seen in NGC4395 is not typical of low luminosity
AGN (Ptak et al 1998).
The normalized excess variance obtained from this observation is 
$\sigma^2_{\rm RMS}\simeq 0.2\pm 0.07$, far above those for 
the sample of Ptak et al (1998).

The present observation of NGC4395 was only half-day long whereas
typical observing time for the Seyfert galaxies in Nandra et al
(1997) and dwarf AGN in Ptak et al (1998) is one-day.
Correction for the observation length makes the $\sigma^2_{\rm RMS}$ of
NGC4395 larger by a factor of 1.4--2, depending on the power spectrum
assumed (e.g., Lawrence \& Papadakis 1993; Nandra et al 1997).
Even without the correction, 
NGC4395 lies close to the extrapolation of the power-law
fit to the $\sigma^2_{\rm RMS}$-$L_{\rm 2-10 keV}$ relation 
(power-law index of $\approx -0.7$) for
their Seyfert-1 sample (Fig. 12). Although $\sigma^2_{\rm RMS}$ for NGC4395
is slightly below the extrapolation, it could be consistent,
given the uncertainty in extrapolating over five orders of magnitude.
A flatter correlation $\sigma_{\rm RMS}^2\propto L_{\rm 2-10 keV}^{-0.5}$
fits the sample of Nandra et al (1997) and NGC4395.


\begin{figure}
\centerline{\psfig{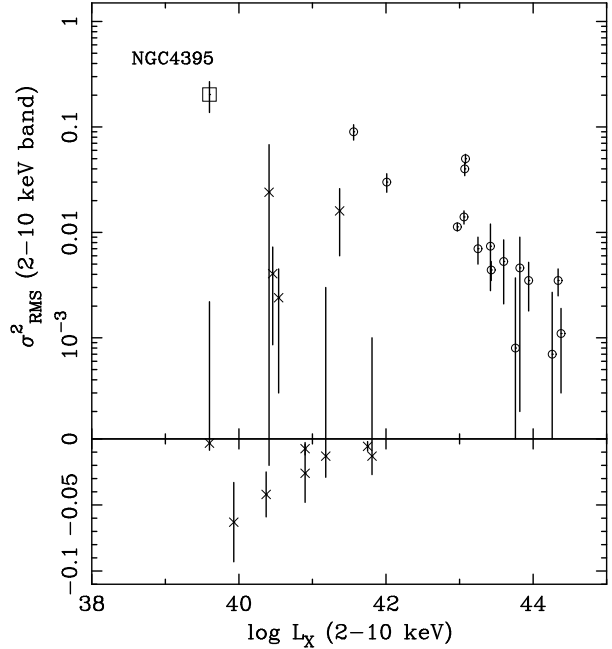}}
\caption{Plot of $\sigma^2_{\rm RMS}$ against X-ray luminosity
for the Seyfert 1 sample (open circles) of Nandra et al (1997), the dwarf AGN
sample (crosses) of Ptak et al (1998) and NGC4395 (open square). Values of the
normalized excess variance were computed for the 2--10 keV data
and the X-ray luminosities are estimated in the 2--10 keV band 
corrected for absorption. The data point for NGC3079 is excluded 
because of the large error bar. Objects with positive $\sigma^2_{\rm RMS}$
are plotted in the upper panel in log scale whilst those with a negative
value are in the lower panel in linear scale.}
\end{figure}

The mass of the black hole in NGC4395 has been estimated to be $\sim
10^5$\Ms\ by various techniques (Lira et al 1999; Kraemer et al 1999;
Filippenko \& Ho 2000). Depending on the technique, it could be as
large as few times $10^6$\Ms.  The most tight constraint is available
from a study of stellar kinematics, which provides the upper limit of
$8\times 10^4$\Ms (Filippenko \& Ho 2000). In any case, the light
black hole mass is consistent with the small bulge of this galaxy
(see the Introduction; e.g., Magorrian et al 1998).  

We have tried to estimate the black hole mass in NGC4395 using the
method employed by Hayashida et al (1998) who used X-ray variability
as an estimator of a black hole mass in scaling that of the Galactic
black hole binary Cyg X-1 whose black hole mass is assumed to be
10\Ms. Note that the black hole masses of the AGNs analysed by
Hayashida et al (1998) are generally one order of magnitude smaller than
those estimated from optical broad line widths, broad-line
reverberation, e.g., Wandel, Peterson \& Malkan (1999). Therefore, this
systematic error should be applied to the mass derived
below. A detailed discussion on the reliability of the method, (i.e.,
the use of high frequency part of the NPSD instead of a ``knee''
frequency which has been found to change in Galactic black hole binary
sources, e.g., Belloni \& Hasinger 1990 for Cyg X-1, as a mass
estimator) is found in Hayashida et al (1998).

Since the quality of the NPSD for NGC4395 is poor, we assume that it
has a broken power-law form, which is canonical for AGN, with power-law
indices obtained for NGC4051 (see Hayashida et al 1998) and that 
the normalization
($N$) follows $1/f$ scaling law. The only free parameter is the break
frequency ($f_b$). We thus describe the NPSD for NGC4395 as
\[ P(f) = \left\{ \begin{array}{ll}
N({\rm N4051})(f_b{\rm (N4051)}/f_b)(f/f_b)^{+0.28} &
 \mbox{($f<f_b$)} \\ 
N({\rm N4051})(f_b{\rm (N4051)}/f_b)(f/f_b)^{-2.01} &
 \mbox{($f\geq f_b$)}
\end{array}
\right. \] 

where $N({\rm N4051})=1.62\times 10^3$ and 
$f_b$(N4051) = $5.69\times 10^{-5}$ Hz
are the normalization and the break frequency for NGC4051. Fitting the
$f_b$ to the NPSD shown in Fig. 3 gives $f_b =
1.82^{+1.64}_{-0.78}\times 10^{-4}$ Hz with $\chi^2 = 4.90$ for 7
degrees of freedom.  The scaling from $M_{\rm BH}$(N4051) (see Fig. 4
in Hayashida et al 1998) then yields $2.8^{+2.4}_{-1.8}\times 10^4$\Ms
for the black hole mass in NGC4395 (the systematic error associated
with this method and the limitation due to our 
assumption of the shape of NPSD should be noted, as mentioned above).

The normalized excess variance of NGC4395 ($\sigma^2_{\rm RMS}\simeq 0.2$)
is larger than NGC4051 ($\sigma^2_{\rm RMS}\simeq 0.1$, Nandra et al 1997)
by a factor of 2.
As noted in Nandra et al (1997) and Papadakis \& Lawrence (1993),
$\sigma^2_{\rm RMS}$ should be proportional to the power at 
a given frequency, if the sampling and power spectra are similar
between the sources.
The power spectrum of NGC4395 could therefore be approximated by 
multiplying the one for NGC4051 by 2.
This leads to a NPSD which agrees with the result obtained above.

Although many assumptions are involved, the black hole mass in NGC4395
implied from the X-ray variability is $\sim 3\times 10^4$\Ms, with
an uncertainty of up to a factor of ten. This is consistent
with the upper limit obtained from the stellar kinematics by
Filippenko \& Ho (1999) and, together with the other estimates, points 
to a relatively light mass black hole. If NGC4395 harbours a black
hole of order of $10^5$\Ms or less, and the scaling law indeed
operates, the low-frequency break, or ``knee'', in the power spectrum,
seen in some Galactic black hole sources and AGN (Belloni \& Hasinger
1990; Hayashida et al 1998; Edelson \& Nandra 1999), to be seen around
$10^{-4}$Hz or higher frequencies as suggested by the fit presented
above (this is however subject to the stability of the break frequency 
at a given black hole mass). 
In order to test this prediction, a longer continuous
observations is needed. However, unlike higher luminosity AGN, a
modest length (say, one-day) observation will suffice. The
rapid X-ray variability together with accompanying spectral change
makes NGC4395 an ideal target for XMM.

The Eddington luminosity for a $10^5M_5$\Ms\ black hole is 
$L_{\rm Edd}\simeq 1.3\times 10^{43}M_5$ \ergps.
The ratio of the 2--10 keV and bolometric luminosities is typically a
few per cent for QSOs and $\sim 10$ per cent for Seyfert galaxies
(e.g., Elvis et al 1994). The bolometric luminosity of NGC4395 may
therefore be in the range of $10^{40}$--$10^{41}$\ergps. 
The Eddington ratio is thus $L_{\rm Bol}/L_{\rm Edd} = 6\times 10^{-3}
M_5^{-1}(f_{\rm HX\rightarrow Bol}/20)$, where $f_{\rm HX\rightarrow Bol}$ 
is the bolometric correction
factor for the 2--10 keV luminosity.
If $M_5$ is about unity or smaller, the Eddington ratio would
be in a range of ordinary Seyfert 1 nuclei.
On the other hand, if $M_5$ is significantly larger than 10, the ratio
would be in the ADAF range (e.g., Kato, Fukue \& Mineshige 1998). 
Since the observed rapid X-ray variability does not fit the ADAF model,
a light black hole ($M_5<1$) is favoured.

The apparent small black hole mass of NGC4395 suggests that 
X-ray variability in AGN correlates with black hole mass,
not directly with luminosity. This may argue that most dwarf AGN
like the sample galaxies in Ptak et al (1998) have massive black holes with low
accretion rates.

\section*{Acknowledgements}

We thank the ASCA team for their efforts on operation of the satellite,
and the calibration and maintenance of the software.
The ROSAT data were obtained through the High Energy
Astrophysics Science Archive Research Center (HEASARC), provided by 
NASA's Goddard Space Flight Center.
ACF and KI thank the Royal Society and PPARC for support, respectively.


\begin{thebibliography}{}

\bibitem{} Barth A.J., Filippenko A.V., Moran E.C., 1999, ApJ, 525, 673

\bibitem{} Belloni T., Hasinger G., 1990, A\&A, 227, L33 

\bibitem{} Condon J.J., Cotton W.D., Greisen E.W., Yin Q.F., Perley R.A., Taylor G.B., Broderick J.J., 1998, AJ, 115, 1693 

\bibitem{} Dickey J.M., Lockman F.J., 1990, ARAA, 1990, 28, 215

\bibitem{} Done C., Mulchaey J.S., Mushotzky R.F., Arnaud K.A., 1992, ApJ, 395, 275 

\bibitem{} Edelson R., Nandra K., 1999, ApJ, 514, 682

\bibitem{} Elvis M., et al, 1994, ApJS, 95, 1

\bibitem{} Fabian A.C., Rees, M.J., 1995, MNRAS, 277, L55

\bibitem{} Filippenko A.V., Sargent W.L.W., 1989, ApJ, 342, L11

\bibitem{} Filippenko A.V., Ho L.C., Sargent W.L.W., 1993, ApJ, 410, L75

\bibitem{} Filippenko A.V., Ho L.C., 2000, ApJ, submitted

\bibitem{} George I.M., Turner T.J., Netzer H., Nandra K., Mushotzky R.F., Yaqoob T., 1998, ApJS, 114, 73
 
\bibitem{} Griffiths R.E., Padovani P., 1990, ApJ, 360, 483

\bibitem{} Guainazzi M., et al, 1998, MNRAS, 301, L1

\bibitem{} Hayashida K., Miyamoto S., Kitamoto S., Negoro H., 1998, ApJ, 500, 642

\bibitem{} Heckman T.M., 1980, A\& A, 87, 152

\bibitem{} Heckman T.M., Armus L., Miley G.K., 1990, ApJS, 74, 833

\bibitem{} Ho L., Filippenko A.V., Sargent W.L.W., 1997a, ApJS, 112, 315

\bibitem{} Ho L., Filippenko A.V., Sargent W.L.W., 1997b, ApJS, 112, 391

\bibitem{} Iwasawa K., 1999, MNRAS, 302, 96

\bibitem{} Kato S., Fukue J., Mineshige S., 1998, Black Hole Accretion Disks, Kyoto: Kyoto University Press

\bibitem{} Kormendy J., Richstone D., 1995, ARAA, 33, 581 

\bibitem{} Kraemer S.B., Ho L.C., Crenshaw D.M., Shields J.C., Filippenko A.V., 1999, ApJ, 520, 564

\bibitem{} Lawrence A., Papadakis I.E., 1993, ApJ, 414, L85

\bibitem{} Lira P., Lawrence A., O'Brien P., Johnson R.A., Terlevich R., Bannister N., 1999, MNRAS, 305, 109

\bibitem{} Lomb N.R., 1976, Ap\&SS, 39, 447


\bibitem{} Magorrian, J. , et al. 1998, AJ, 115, 2285 

\bibitem{} Morales R., Fabian A.C., Reynolds C.S., 2000, MNRAS, in press

\bibitem{} Moran E.C., Filippenko A.V., Ho L.C., Shields J.C., Belloni T., Comastri A., Snowden S.L., Sramek R.A., 1999, PASP, 111, 801

\bibitem{} Morrison R., McCammon D., 1983, ApJ, 270, 119

\bibitem{} Nandra K., George I.M., Mushotzky R.F., Turner T.J., Yaqoob T., 1997, ApJ, 476, 70 

\bibitem{} Otani C., et al, 1996, PASJ, 48, 211

\bibitem{} Ptak A., Yaqoob T., Mushotzky R., Serlemitsos, P., Griffiths R., 1998, ApJ, 501, L37 

\bibitem{} Radecke H.-D., 1997, A\&A, 319, 18

\bibitem{} Reynolds C.S., 1997, MNRAS, 286, 513

\bibitem{} Richstone D., et al, 1998, Nat, 395, 14

\bibitem{} Serlemitsos P.J., et al, 1995, PASJ, 47, 105

\bibitem{} Terashima Y., 1998, PhD thesis, Nagoya University, Japan

\bibitem{} Turner T.J., George I.M., Nandra K., Turcan D., 1999, ApJ, 524, 667

\bibitem{} Wandel A., Peterson B.M., Malkan M.A., 1999, ApJ, 526, 579

\bibitem{} Ward M.J., Done C., Fabian A.C., Tennant A.F., Shafer R.A., 1988, ApJ, 324, 767 

\end{thebibliography}
\end{document}